\documentclass{PoS}
\usepackage{subfig}

\title{Associated production of top quarks with the Higgs boson at $\sqrt{s}=13$ TeV}

\ShortTitle{ttH and tH production at $\sqrt{s}=13$ TeV}

\author{\speaker{Georgios Konstantinos Krintiras} on behalf of the CMS Collaboration\\
        Universit\'e catholique de Louvain, Louvain-la-Neuve, Belgium\\
        E-mail: \email{gkrintir@cern.ch}}

\abstract{The top quark, being the heaviest elementary fermion known in the Standard model, has the largest coupling to the Higgs boson. 
The associated production of top quarks with the Higgs boson, either in pairs (t$\bar{\rm{t}}$H) or singly (tH), provides direct experimental access to the top-Higgs coupling $y_{\rm{t}}$.
The t$\bar{\rm{t}}$H (tH) production mode, while proceeding at a rate of about 100 (1000) times smaller than gluon fusion, 
bears a highly distinctive experimental signature, which includes leptons and/or jets from the decay of the two (single) top quarks.
The latest results of ttH searches at a center-of-mass energy of 13 TeV corresponding to an integrated luminosity of up to $35.9\ \rm{fb}^{-1}$ as collected from CMS are shown and tantalizing 
evidence is found for measuring this crucial process with sufficient 
precision. However, higher precision data set is needed in order to confirm or disprove the previous observed excess. 
Initial searches for tH production mode at a center-of-mass energy of 13 TeV achieve comparable sensitivity to that of the Run 1 analysis. 
}

\FullConference{XXV International Workshop on Deep-Inelastic Scattering and Related Subjects\\
		3-7 April 2017\\
		University of Birmingham, UK}

\begin{document}

\section{Introduction}
The fermion masses originate entirely from the interactions manifested in the Yukawa sector of the Standard Model (SM),
which accounts for nearly two third from a total of 19 free SM parameters. Due to the large quark mass, 
the Yukawa coupling of the top quark with the Higgs boson ($y_{\rm{t}}$) is almost unity, 
and along with the apparent near criticality of the Higgs potential parameters \cite{cite_key1}, hint to 
a special role of the top-Higgs Yukawa interaction in the mechanism of electroweak symmetry breaking (EWSB). Combined ATLAS and CMS results 
based on the LHC Run 1 data set on the associated production of a top quark pair with the Higgs boson (ttH) --  a direct probe of $y_{\rm{t}}$-- 
 exhibited an intriguing excess: the measured rate was above the SM prediction with a statistical significance corresponding to $2.3\sigma$ \cite{cite_key2}.

With the increase of the LHC energy from 8 to 13 TeV for Run 2, the ttH production cross section is expected to increase by a factor four. 
However such process still remains rare and searches for ttH production have been driven by the higher sensitivity achieved in Higgs decay modes with larger branching fraction i.e.,
$\rm{H} \rightarrow \rm{bb}$, $\rm{H} \rightarrow \rm{WW}$ and $\rm{H} \rightarrow \rm{\tau\tau}$. 
Clean final states containing two photons or four leptons, though currently limited for the expected signal yields in these modes are just a few events,
remain promising candidates for the first observation of ttH production.
The latest results of ttH searches at a center-of-mass energy of 13 TeV corresponding to an integrated luminosity of up to $35.9\ \rm{fb}^{-1}$ as measured in CMS \cite{cms} are presented. 

The associated production of a single top quark with the Higgs boson (tH) is also reviewed due to its unique Born-level sensitivity on the relative sign of 
the $y_{\rm{t}}$ coupling with respect to the coupling to gauge bosons. For instance, an
opposite i.e., inverted, sign in the Yukawa coupling with respect to the SM prediction would jeopardize perturbative
unitarity at a high energy scale and in turn would lead to a striking visible enhancement of the tH cross section at
the EWSB scale.

\section{Associated production of a top quark pair with the Higgs boson}
\subsection{$\rm{t}\bar{\rm{t}}$($\rm{H}\rightarrow \rm{WW}$, $\rm{\tau\tau}$, $\rm{ZZ^{*}}$) final states}
The targeted events for these Higgs decays with at least one of the top quarks decaying leptonically contain two leptons, either electrons
or muons, of the same charge ($2l\rm{ss}$), or more than three leptons ($\geq3l$), and hadronic jets compatible with the hadronization of b quarks (b ``tagged'' jets). 
In order to gain sensitivity of the analysis, an additional categorization is performed by the lepton flavor, the presence of 2 b tagged jets -- with b ``tight''  representing the
category that has two b tagged jets in the final state and the complementary b ``loose'' category -- and the total lepton charge. 
A multivariate (MVA) method for lepton identification is employed to identify ``prompt'' leptons i.e., leptons originating from $\rm{t}\bar{\rm{t}}$ and Higgs decays.
Dominant backgrounds include reducible processes 
that are responsible for jets being misidentified as prompt leptons, e.g. processes involving  t$\bar{\rm{t}}+\rm{jets}$, and irreducible contribution mainly from the associated production 
of a top quark pair with a vector boson (t$\bar{\rm{t}}\rm{V}$).

\begin{figure}[htp]
\centering
\subfloat[][]{\includegraphics[width=0.32\textwidth]{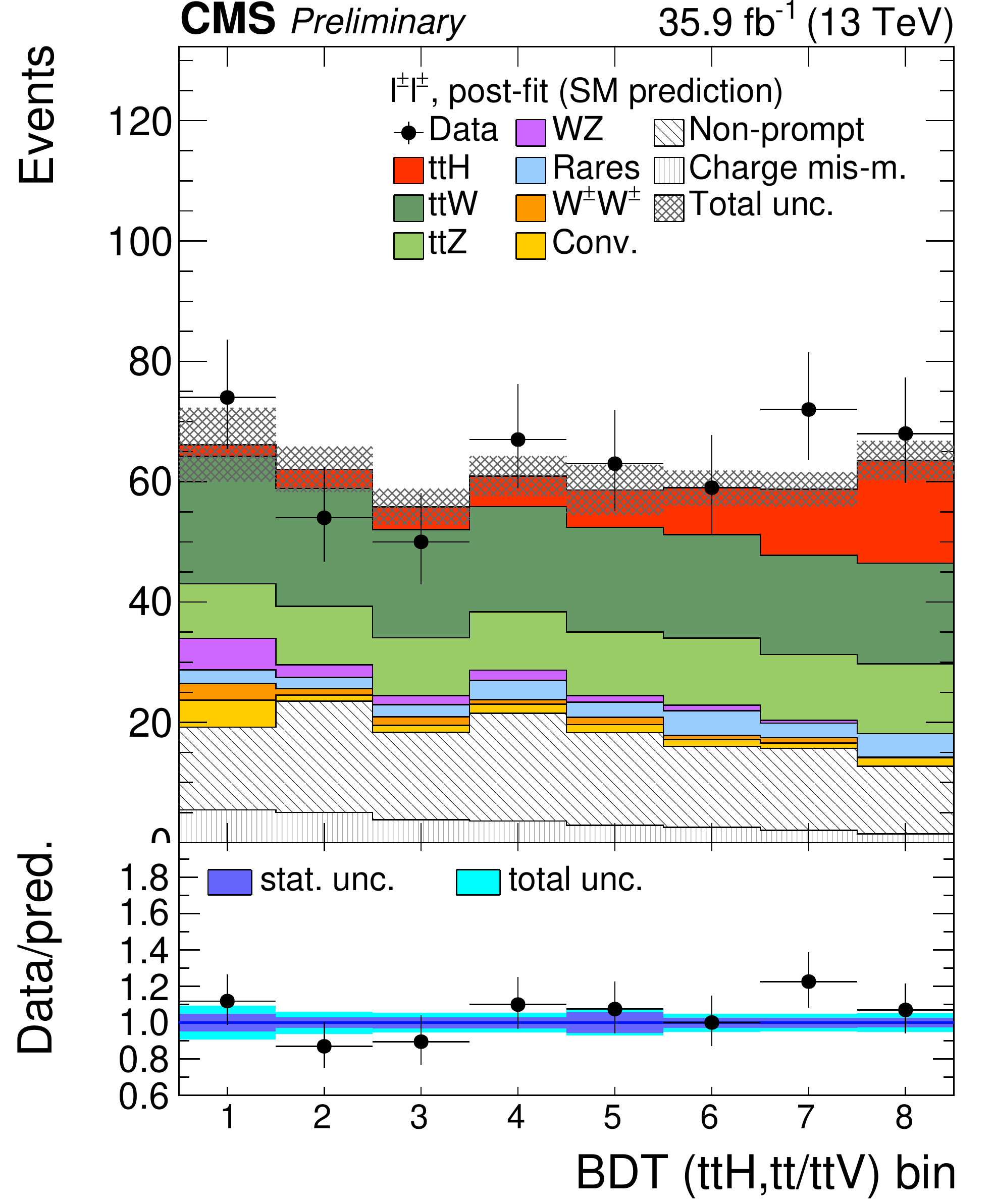}\label{bdt_2lssl}}
\subfloat[][]{\includegraphics[width=0.32\textwidth]{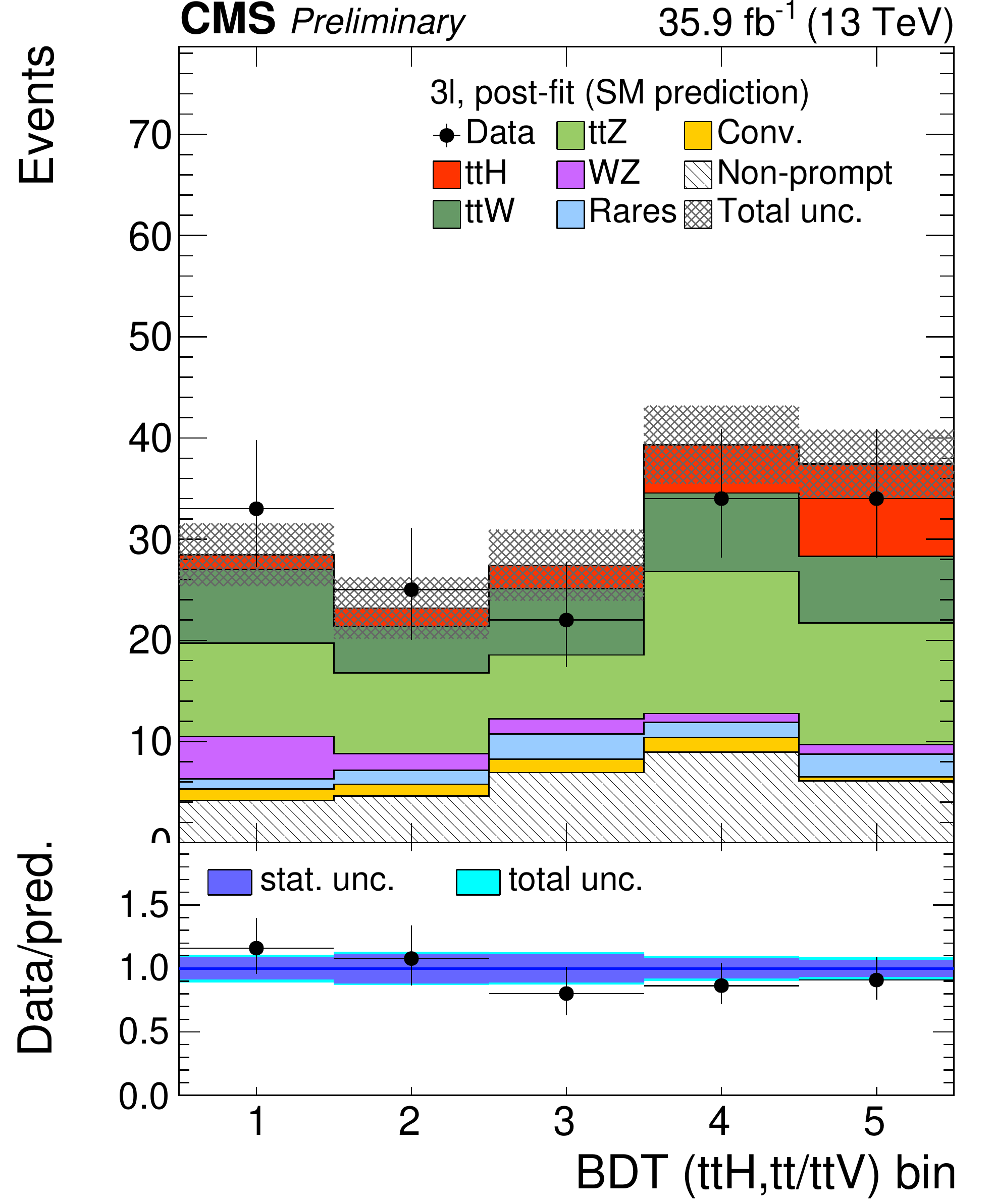}\label{bdt_3l}}
\subfloat[][]{\includegraphics[width=0.32\textwidth]{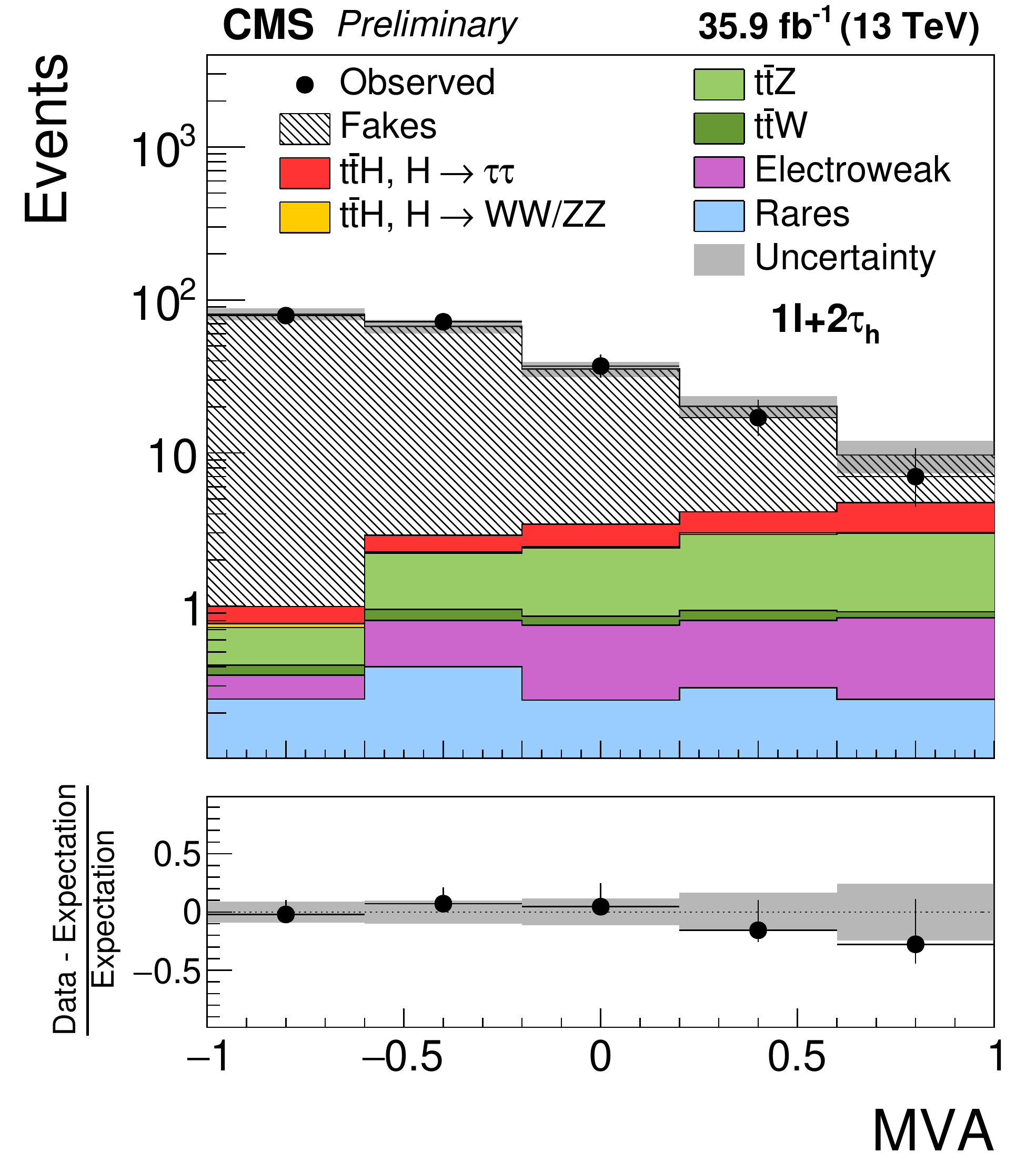}\label{mva_1l2tau}}\\
\caption{
Combination of the BDT classifier outputs in the bins used for signal extraction in the $2l\rm{ss}$ (a) and $3l$ (b) channel, respectively  \cite{cite_key3}. 
Distribution in the discriminating observable used for the signal extraction in the $2l+2\tau_{h}$ category \cite{cite_key4} (c). }
\label{fig:vtxXYFits_Fills5527And5563}
\end{figure}

Signal extraction is performed using final discriminators obtained with a combined MVA approach. The
MVA outputs, separately trained against t$\bar{\rm{t}}$ and t$\bar{\rm{t}}\rm{V}$ scenarios are used to build two boosted decision tree (BDT) discriminators,
which are further split into bins of a roughly constant number of background events and an increasing number of signal events (Fig.\ref{bdt_2lssl}-\ref{bdt_3l}).
The MVA takes into account several kinematic (such as the maximum pseudorapidity of the two ``leading'' i.e., highest $p_{T}$, leptons) and 
angular (such as the angular separation between the leptons and the jets) variables, and the jet multiplicity. 
Additional variables are considered in the $2l\rm{ss}$ and $3l$ event categories including BDT scores for reconstructing hadronic top decays and
matrix element (MEM) weights for discriminating signal against irreducible backgrounds, respectively. 
Dedicated identification discriminants for $\tau_{h}$ decays to hadrons have been developed and complemented by means of two different MEM and BDT techniques 
separately in three final states i.e., $2l+2\tau_{h}$, $2l\rm{ss}+1\tau_{h}$ and $3l+1\tau_{h}$ (Fig.\ref{mva_1l2tau}).

\subsection{$\rm{t}\bar{\rm{t}}$($\rm{H}\rightarrow \rm{bb}$) final states}

\begin{figure}[htp]
\centering
\subfloat[][]{\includegraphics[width=0.32\textwidth]{check.pdf}\label{bdt_6j4b}}
\subfloat[][]{\includegraphics[width=0.32\textwidth]{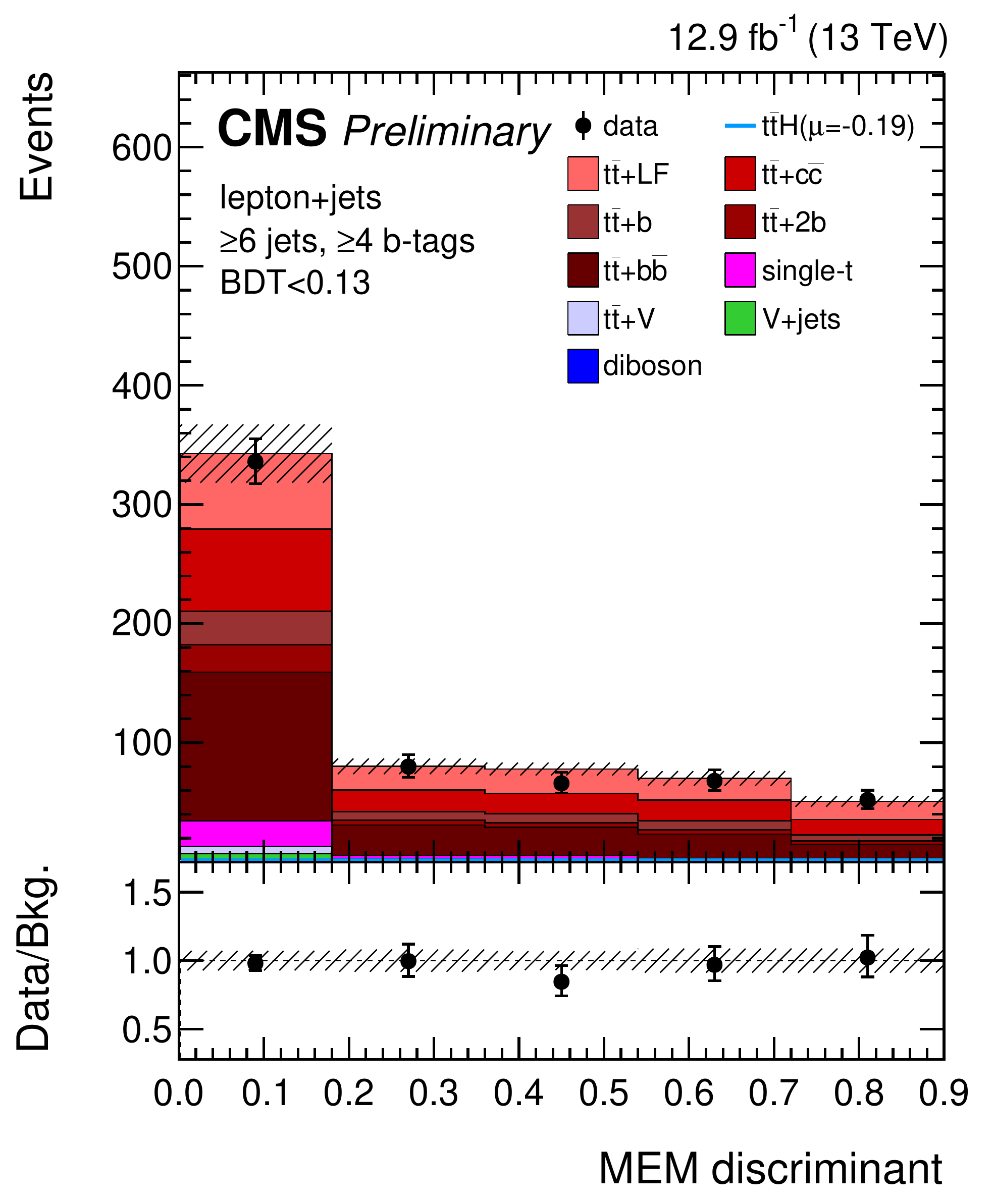}\label{mem_lowbdt_6j4b}}
\subfloat[][]{\includegraphics[width=0.32\textwidth]{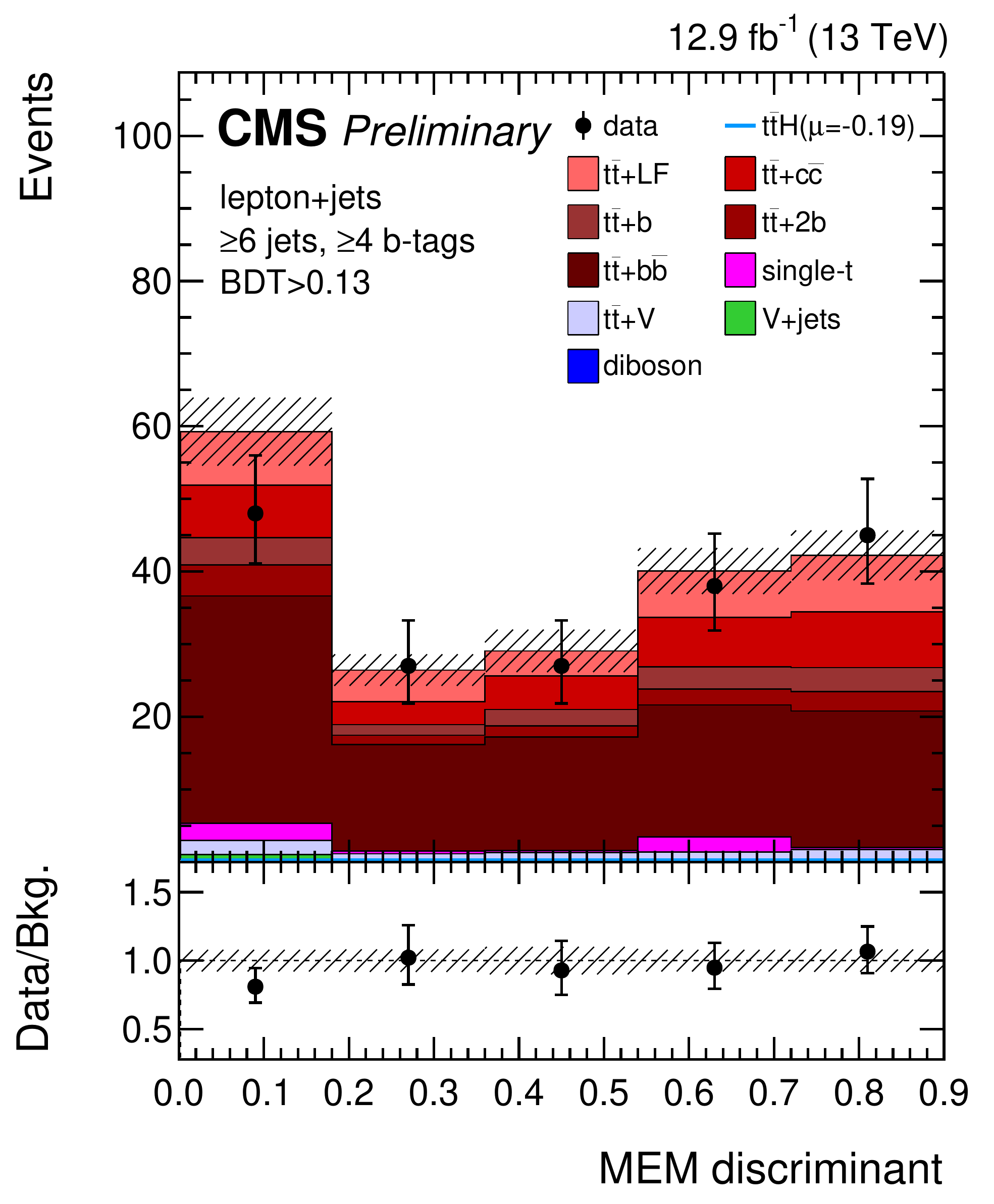}\label{mem_highbdt_6j4b}}\\
\caption{
BDT output distribution in the $\geq6$jets,$\geq4$b-tags category including semi-leptonic top quark decays (a).
Final discriminant shapes (MEM) in the same analysis category with low ($\rm{BDT<0.13}$, b) and high ($\rm{BDT>0.13}$, c) BDT output  \cite{cite_key6}. 
}
\label{fig:vtxXYFits_Fills5527And5563}
\end{figure}

The strategy of the $\rm{t}\bar{\rm{t}}$($\rm{H}\rightarrow \rm{bb}$) analysis with the top quarks decaying 
either leptonically or semi-leptonically is defined by categorizing the
selected events based on the reconstructed jet and b tagged jet multiplicity into distinct categories (``number of jets, number of b-tags'') with different background composition.
Discriminators for each of the categories are constructed with a two-tier MVA approach, which is found to improve the predictions from theory and
to reduce the uncertainties from a combined fit over the final discriminant output; first, the BDT algorithm combines event kinematic properties (such as the invariant masses and angular
correlations of combinations of jets and leptons, Fig.\ref{bdt_6j4b}); second, according to the BDT discriminant output two sub-categories are produced 
that differ in the composition of the hadronically decaying background components; 
finally, the MEM technique is used as the final discriminant in each resulted sub-category (Fig.\ref{mem_lowbdt_6j4b}-\ref{mem_highbdt_6j4b}).

\subsection{$\rm{t}\bar{\rm{t}}$($\rm{H}\rightarrow \rm{\gamma\gamma}$) final states}

\begin{figure}[htp]
\centering
\subfloat[][]{\includegraphics[width=0.32\textwidth]{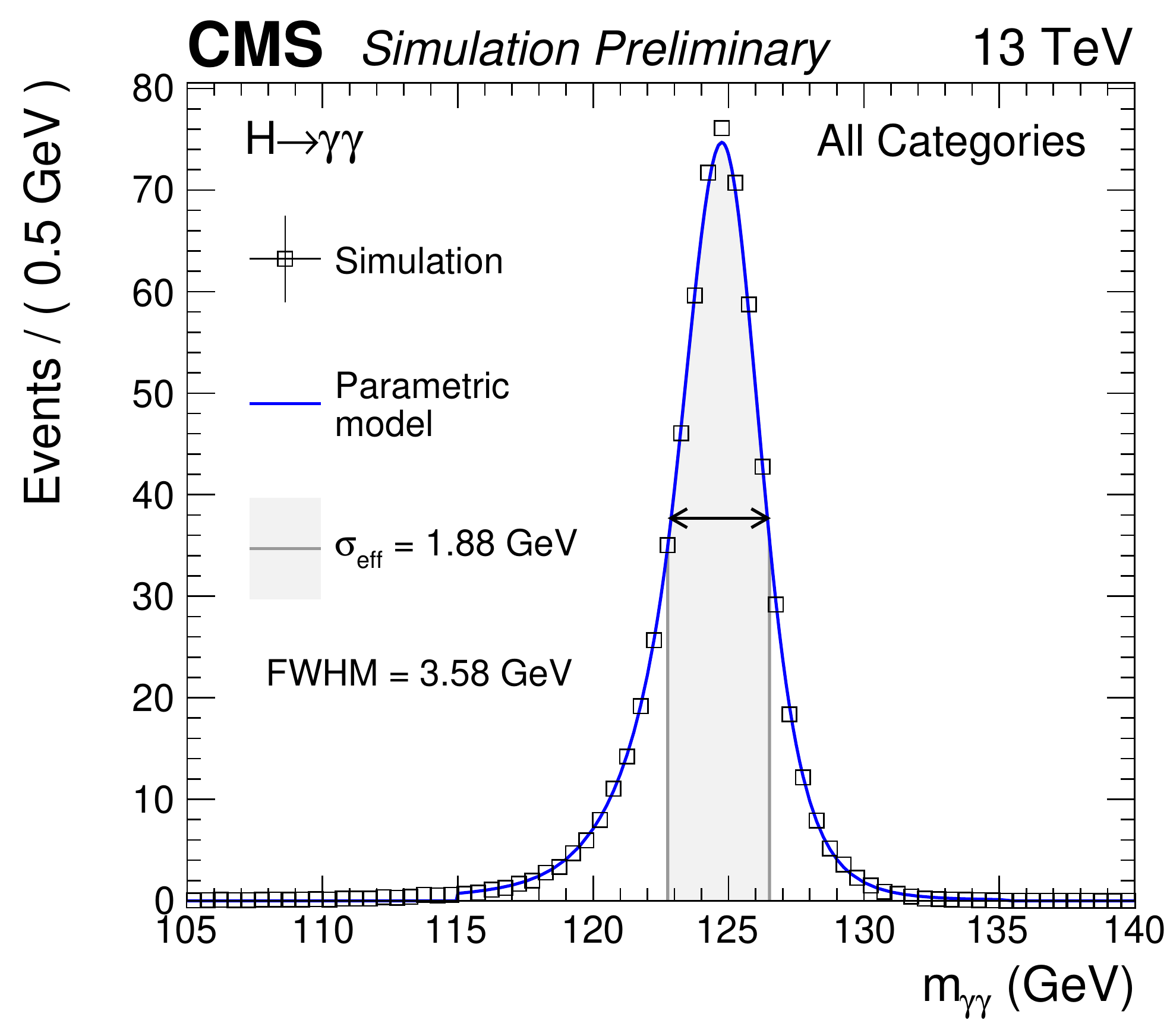}\label{mass_res_allcats}}
\subfloat[][]{\includegraphics[width=0.32\textwidth]{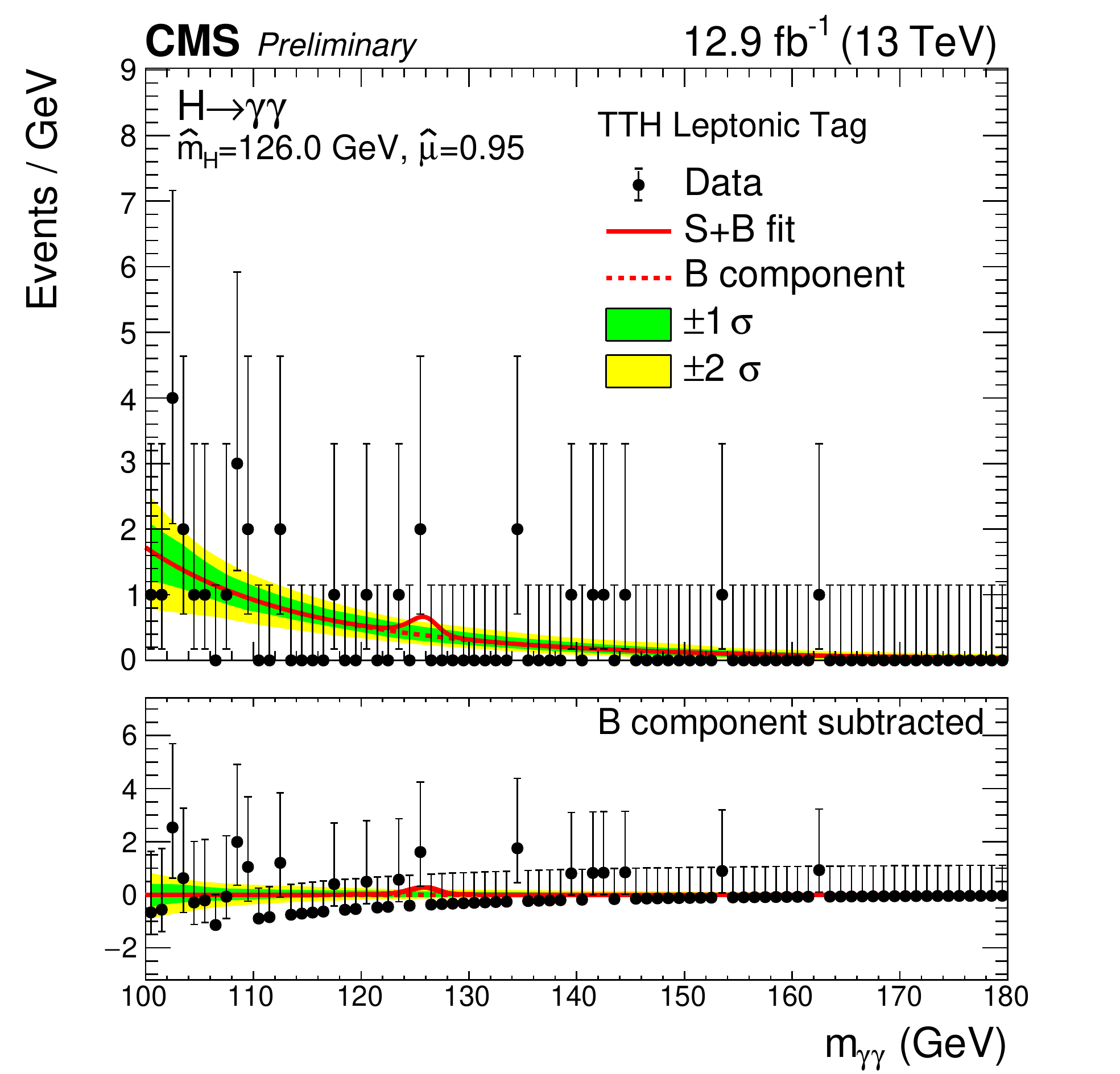}\label{tth_leptonic}}
\subfloat[][]{\includegraphics[width=0.32\textwidth]{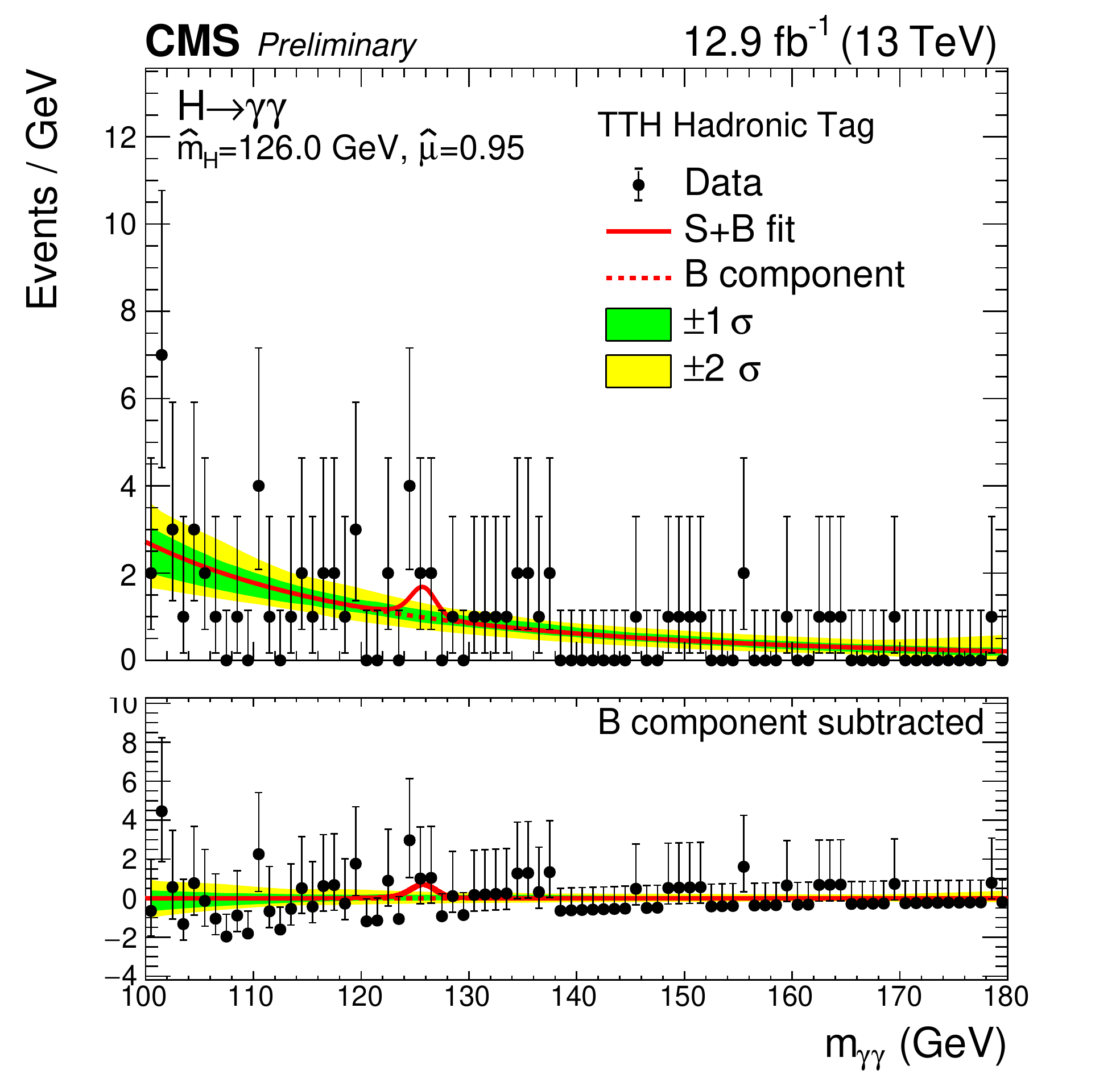}\label{tth_hadronic}}\\
\caption{
Parametrized signal shape for all event categories combined together for a simulated $\rm{H}\rightarrow \rm{\gamma\gamma}$ sample (a).
Data points (black) and signal plus background model fits in ttH Leptonic (b) and ttH Hadronic (c) categories \cite{cite_key7}. 
The bottom plots show the residuals after background subtraction. 
An updated analysis with a total luminosity of  $35.9\ \rm{fb}^{-1}$ has been presented in \cite{tth_gammagamma_up}.  
}
\label{fig:vtxXYFits_Fills5527And5563}
\end{figure}

This analysis provides clean final state signatures since it exploits the advantage 
to measure diphoton mass with excellent resolution (Fig.\ref{mass_res_allcats}).
To improve the sensitivity of the analysis, events are classified targeting (``tagging'') different production
mechanisms according to their mass resolution and predicted signal-to-background ratio.
Events produced in association with a top quark pair feature two b quarks from the decay of
the top quarks, and may be accompanied by charged leptons (``ttH Leptonic'') and/or additional jets (``ttH Hadronic''). 
Signal is extracted using a parametrized model of the Higgs boson mass shape obtained from
simulation (Fig.\ref{mass_res_allcats}), while the combined signal and background models for these two categories are shown
on Fig.\ref{tth_leptonic}-\ref{tth_hadronic}. The model used to describe the background is extracted from data with the discrete
profiling method, which is designed to estimate the systematic uncertainty associated with choosing
a particular analytic function to fit the background distribution. The method treats the choice of the
background function as a discrete parameter in the likelihood fit to the data. The resulting systematic
uncertainty is then calculated in an analogous way to systematic uncertainties associated with other
contributions. 

\section{Associated production of a single top quark with the Higgs boson}

\begin{figure}[htp]
\centering
\subfloat[][]{\includegraphics[width=0.32\textwidth]{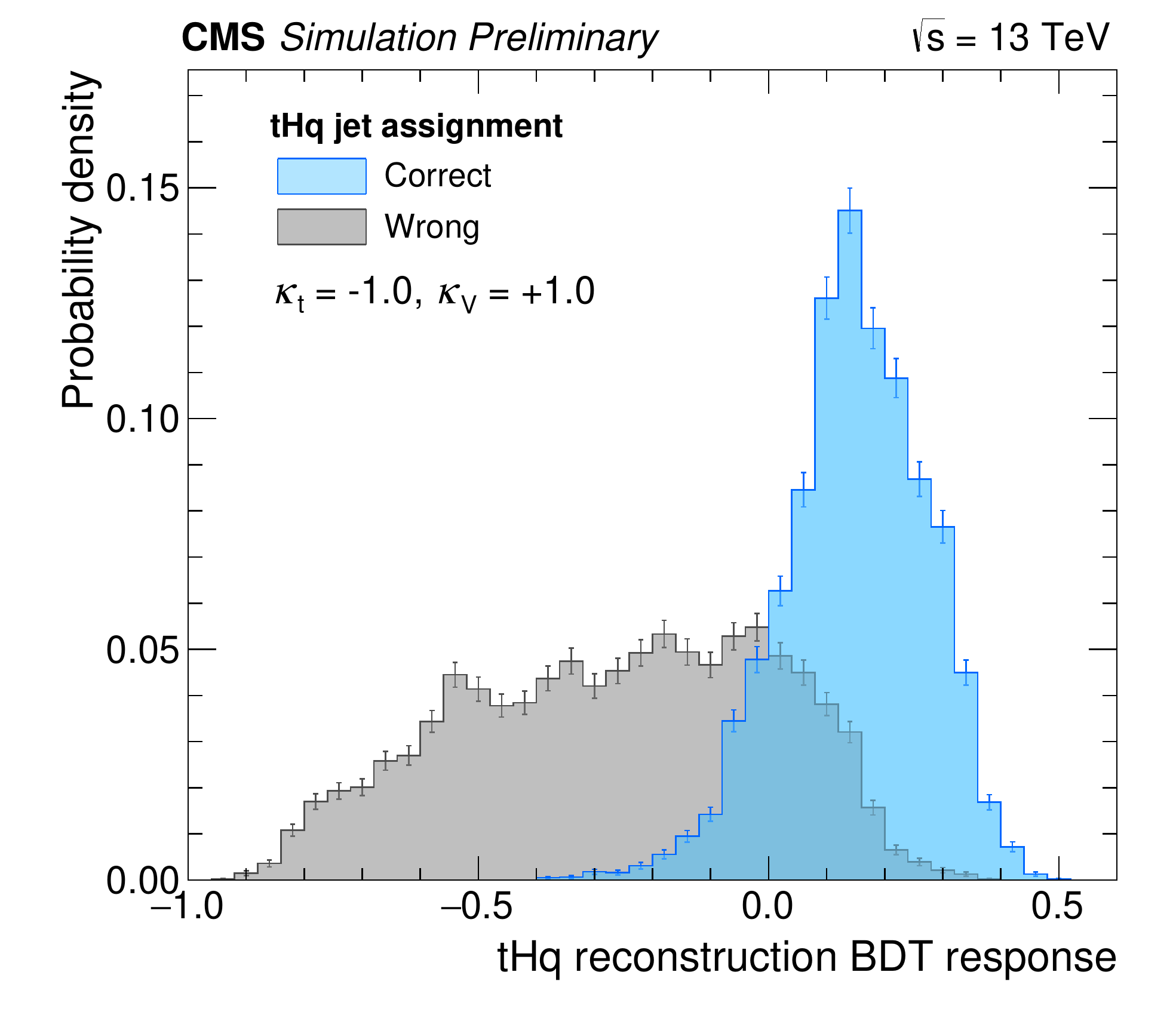}\label{bdt_thq}}
\subfloat[][]{\includegraphics[width=0.32\textwidth]{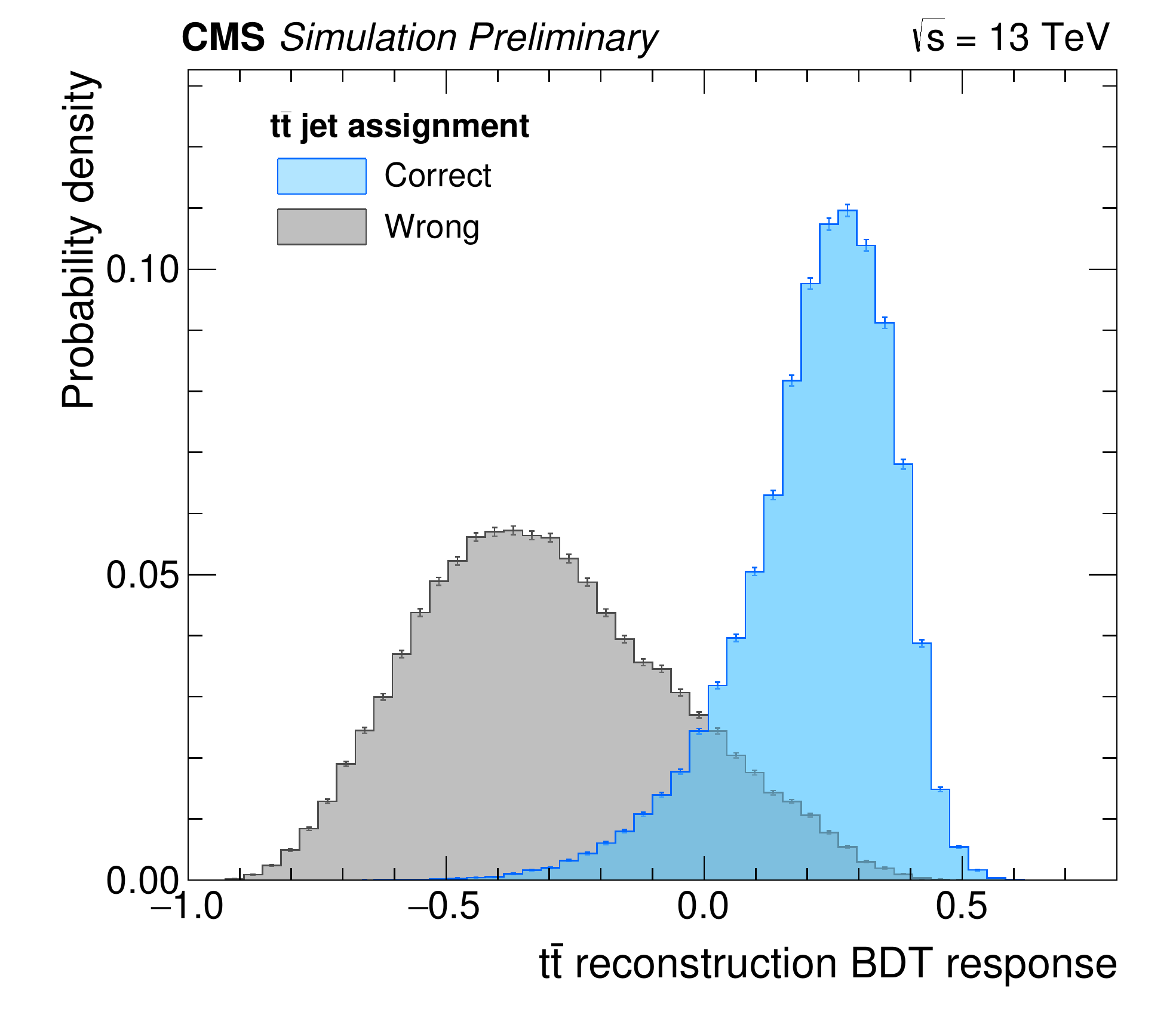}\label{bdt_tt}}
\subfloat[][]{\includegraphics[width=0.32\textwidth]{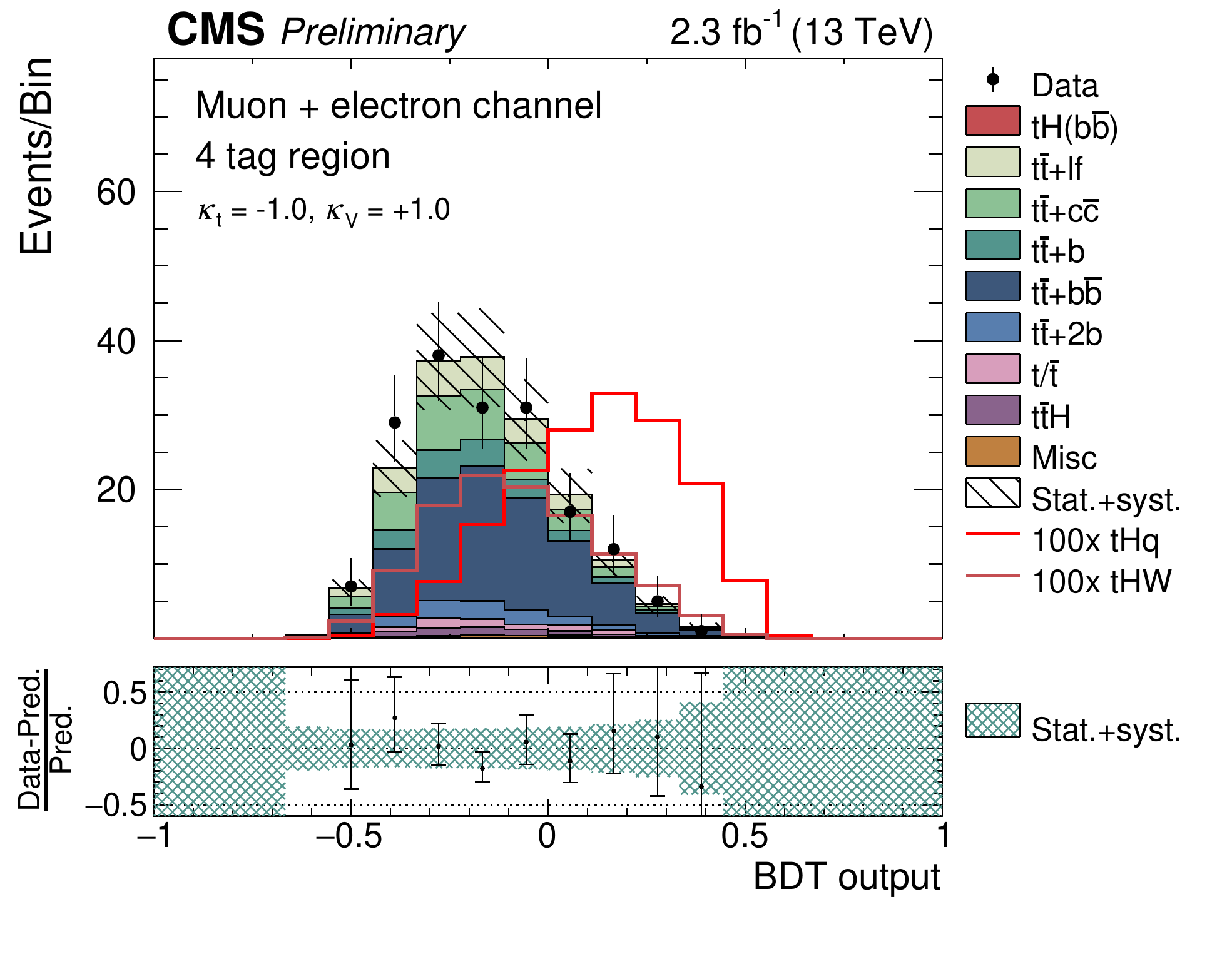}\label{bdt_final}}\\
\caption{
Response of the signal (a) and $\rm{t}\bar{\rm{t}}$ (b) reconstruction classifiers (BDT). Final BDT response in events with exactly four b-tagged
jets \cite{cite_key8}. 
}
\label{fig:vtxXYFits_Fills5527And5563}
\end{figure}

A direct search for t($\rm{H}\rightarrow \rm{b\bar{b}}$) production has been performed at a center-of-mass energy of 13 TeV including semi-leptonic top
quark decays. Two types of reconstructions,
with one set representing intrinsic properties  of  signal (Fig.\ref{bdt_thq}) and  the  other  set  consisting  of  variables  characteristic  for 
the dominant $\rm{t}\bar{\rm{t}}$ background (Fig.\ref{bdt_tt}), form  together with the lepton charge (global variable) the list of input variables for the final BDT (Fig.\ref{bdt_final}).
 For the first time, the upper asymptotic $\rm{CL_{\rm{S}}}$ limits are reported separately for 51 coupling configurations by evaluating the corresponding BDT response (Fig.\ref{thq_bb}). 

\section{Summary}

\begin{figure}[htp]
\centering
\subfloat[][]{\includegraphics[width=0.42\textwidth]{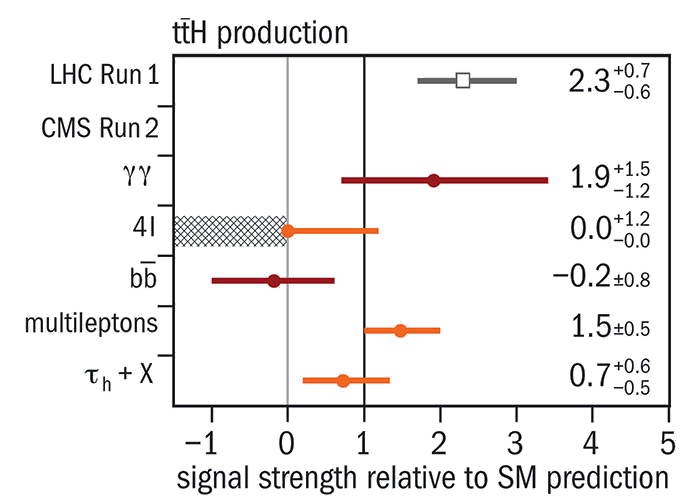}\label{tth}}
\subfloat[][]{\includegraphics[width=0.42\textwidth]{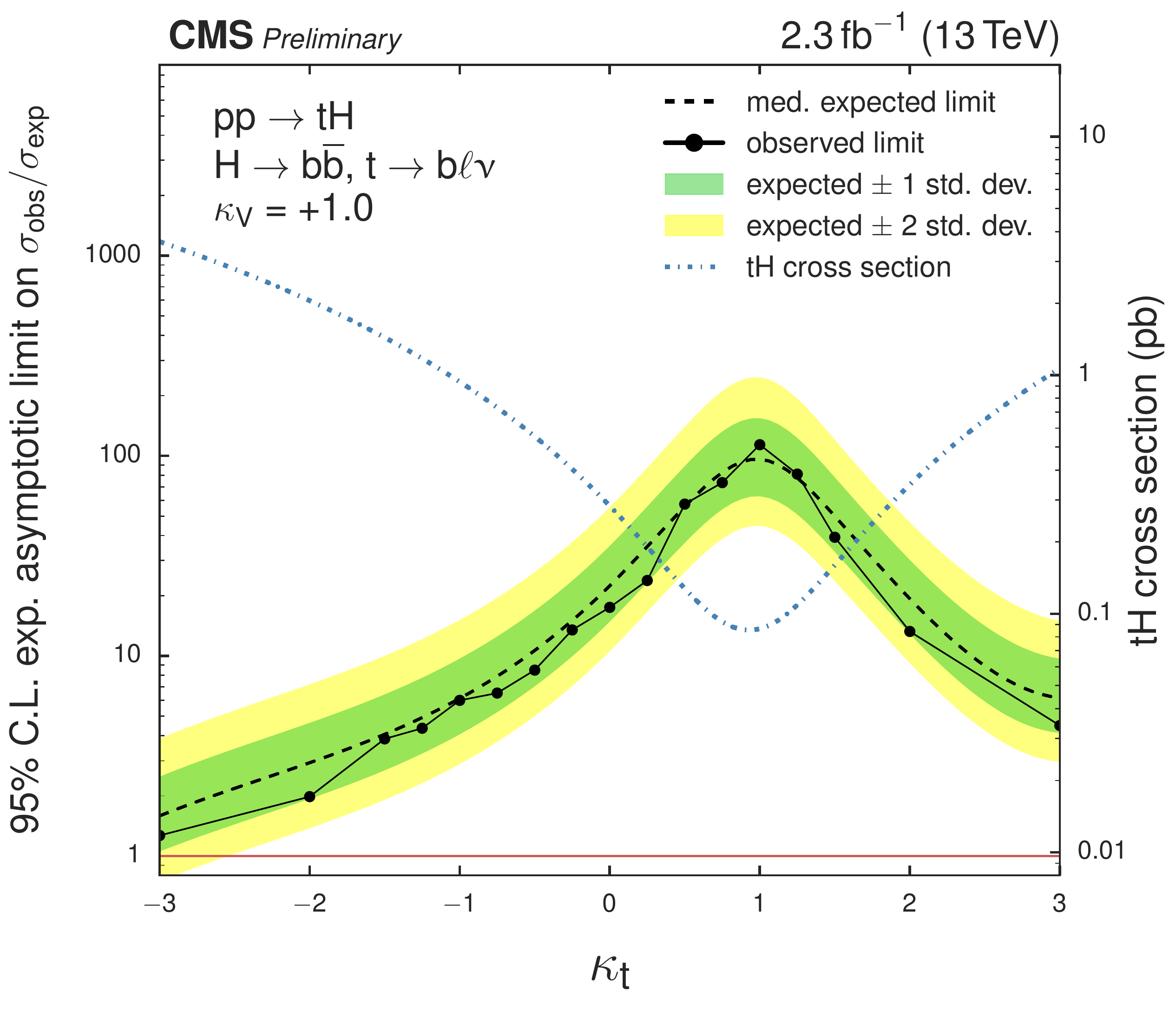}\label{thq_bb}}\\
\caption{
The measured signal strength $\mu=\frac{\sigma}{\sigma_{\rm{SM}}}$ for each of the considered Run 2 analysis
\cite{cite_key3,cite_key4,cite_key5,cite_key6,cite_key7} along with LHC Run 1 combination \cite{cite_key2}.
Plot was adapted from \cite{cite_key10} (a). 
Upper limits on tH scenario with an inverted $y_{\rm{t}}$ coupling with respect to the coupling to gauge bosons (b).  
For illustration, also the tH cross sections are given on the right $y$ axis  \cite{cite_key8}. }
\label{fig:vtxXYFits_Fills5527And5563}
\end{figure}

Direct experimental access to the top-Higgs coupling comes from the study of the associated production of top quarks with the Higgs boson.
The latest results of ttH searches at CMS (Fig.\ref{tth}) show a tantalizing evidence of measuring this crucial process with 
sufficient precision to confirm or disprove the previous observed excess. The sensitivity is currently driven by final states containing multiple leptons,
a compromise between expected signal yield and background uncertainty.

The observed (expected)
limit on the cross section for the production of the Higgs boson in association with a
single top quark in the scenario with an inverted $y_{\rm{t}}$ coupling with respect to the coupling to gauge bosons is 6.0 (6.4) times the predicted value. 
The sensitivity of this analysis is already comparable to that of the Run 1 search \cite{cite_key9}.

With a larger data set it should be possible to have clear evidence for ttH and for anomalous, if proven, tH production by the end of Run 2.

\end{document}